\documentclass[10pt, conference]{IEEEtran}
\IEEEoverridecommandlockouts
\usepackage{bm}
\usepackage{comment}
\usepackage{array}
\usepackage{multirow}
\usepackage{color}
\usepackage[utf8]{inputenc}
\usepackage{graphicx}
\graphicspath{ {images/} }

\usepackage{cite}
\usepackage{amsmath,amssymb,amsfonts}
\usepackage{algorithmic}
\usepackage{textcomp}
\def\BibTeX{{\rm B\kern-.05em{\sc i\kern-.025em b}\kern-.08em
    T\kern-.1667em\lower.7ex\hbox{E}\kern-.125emX}}
\begin{document}

\title{Automatic Device Classification from Network Traffic Streams of Internet of Things\\
}

\author{
\IEEEauthorblockN{Lei Bai, Lina Yao, Salil S. Kanhere}
\IEEEauthorblockA{
\textit{University of New South Wales}\\
Sydney, Australia \\
baisanshi@gmail.com, \{lina.yao, salil.kanhere\}@unsw.edu.au}
\and
\IEEEauthorblockN{Xianzhi Wang}
\IEEEauthorblockA{
\textit{University of Technology Sydney}\\
Sydney, Australia \\
sandyawang@gmail.com}
\and
\IEEEauthorblockN{Zheng Yang}
\IEEEauthorblockA{
\textit{Tsinghua University}\\
Beijing, China \\
hmilyyz@gmail.com}
}

\maketitle

\begin{abstract}
With the widespread adoption of Internet of Things (IoT), billions of everyday objects are being connected to the Internet. Effective management of these devices to support reliable, secure and high quality applications becomes challenging due to the scale. As one of the key cornerstones of IoT device management, automatic cross-device classification aims to identify the semantic type of a device by analyzing its network traffic. It has the potential to underpin a broad range of novel features such as enhanced security (by imposing the appropriate rules for constraining the communications of certain types of devices) or context-awareness (by the utilization and interoperability of IoT devices and their high-level semantics) of IoT applications.
We propose an automatic IoT device classification method to identify {\em new} and {\em unseen} devices. The method uses the rich information carried by the traffic flows of IoT networks to characterize the attributes of various devices. We first specify a set of discriminating features from raw network traffic flows, and then propose a LSTM-CNN cascade model to automatically identify the semantic type of a device. Our experimental results using a real-world IoT dataset demonstrate that our proposed method is capable of delivering satisfactory performance. We also present interesting insights and discuss the potential extensions and applications. 
\end{abstract}


\section{Introduction}

Internet of Things (IoT) aims to interconnect everyday objects such as cars, fridges, watches, and thermostats, thus facilitating easy collection and exchange of data.
IoT is expected to fundamentally revolutionize our lives and impact a wide range of application domains including healthcare, transportation, energy, and infrastructure. 
The recent rapid development of IoT has resulted in a sharp increase in the number of devices being connected to the Internet. It is reported by Gartner that there will be over 20 billion connected IoT devices by 2020\footnote{https://www.gartner.com}. These IoT devices belong to different categories, such as cameras, televisions, fitness devices, and environmental sensors. They provide various services in all aspects to serve the society, from smart home to smart building and from smart city to smart factory. From the network operation's perspective, to \textbf{automatically} and \textbf{quickly} identify the semantic category of a device (e.g., camera, fitness/medical device, environmental sensor, etc) can be of great value in many ways. For device configuration, network administrators have to configure different rules for each device depending on its type, which is time-consuming, prone to errors and unscalable given the large number of IoT devices typically installed in an enterprise setting. For ensuring network security, administrators may wish to prohibit the use of certain types of devices (e.g. cameras in a secure facility). For QoS guarantee, traffic from different types of devices may be given different priorities, for example, network flows from healthcare devices should have precedence over those from entertainment devices during periods of high load.

\begin{figure*}
\setlength{\abovecaptionskip}{0.cm}
\setlength{\belowcaptionskip}{-0.cm}
\centering
\includegraphics[width = 0.8\linewidth]{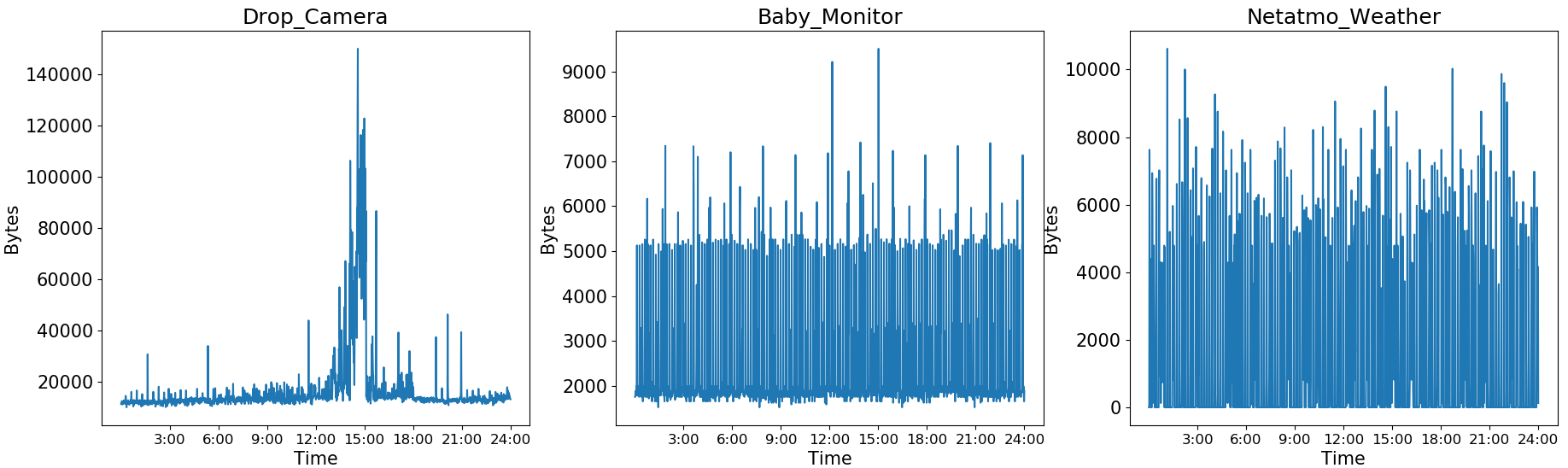}
\caption[]{Daily Network traffic volume from three IoT devices. }
\label{comparision}
\end{figure*}

Therefore, there is an urgent need for automatic categorization of heterogeneous IoT devices to provide reliability, security, and improved QoS to upstream applications. However automatic device classification by analysing network traffic is a non-trivial task due to the dynamic and complex nature of network traffic in IoT. The network traffic of a device may vary a lot at different times associated with user interactions or client-server communications. It is thus hard to characterize device's network traffic into a fix pattern. Moreover, each device category usually contains many different devices with similar functions offered by different device vendors and manufacturers or having the different hardware/firmware versions. It is challenging to build up an invariant profile across different types of devices. For instance, Drop Camera and Withings Smart Baby Monitor are both {\em Cameras} but are produced by different manufactures. However the network traffic pattern for these devices are significantly different by a simple observation of their traffic volume as shown in Figure \ref{comparision}. In fact, as shown in Figure \ref{comparision}, the traffic from a device belonging to a different semantic type, a Netamo Weather station has some resemblance with the traffic of the baby monitor. To achieve accurate classification it is thus important to find shared patterns in the traffic for devices that belong to a category but also exclude similarities between devices that belong to different categories. 





In recent years, automatic classification of IoT devices by examining network traffic has been explored in \cite{second}\cite{third}\cite{eighteen}\cite{nineteen}. However, these studies either only discern devices with same hardware/firmware versions \cite{nineteen} or subject to strong constraints. More specifically, their approaches use the data from the same device for training and testing on a temporal basis \cite{third}\cite{second} and can thus only achieve intra-device identification and classify devices already exist in the training dataset. But in reality, it is impossible to include all IoT devices in the training phase. Thus, automatic cross-device classification is necessary for real world applications.

To fill this gap, we propose {\em unseen device identification} by fully exploiting the implicit correlations and invariant patterns from the underlying IoT network traffic flows. Our goal is to automatically identify new and unseen device types by analyzing IoT network traffic, where we use traffic streams to characterize the devices' semantic categories. 
Our main contributions of this paper can be summarized as follows: 

\begin{itemize}

\item We propose a unified framework for automatic IoT device classification, wherein we design an approach for deriving features and extracting the invariant dependencies across devices. Specially, we provide an in-depth empirical study on the key characteristics of real world IoT traffic data. We also propose a LSTM-CNN cascade model to classify the IoT devices via capturing the global and local temporal correlations in a supervised manner; 

\item[--] To the best of our knowledge, we are the first to leverage time-dependencies of network traffic to automatically classify unknown IoT devices into categories according to their function. Compared to existing works, our approach would be easy to scale up for better practical use given the large number of IoT devices;

\item We evaluate our approach on the real IoT dataset. Our proposed model could achieve satisfactory accuracy on classifying new devices with only a small training set. We also show that our method outperforms a wide range of baseline algorithms. 

\end{itemize}

The rest of this paper is organized as follows. We summarize the related work in section II. Section III presents the proposed approach and technical details. The experimental results are given in section IV. We also include an open discussion about classifying IoT devices via mining network traffic in this part. Finally the paper is concluded in section V.

\section{Related work}
In this section, we present an overview of closely related work which is grouped under three broad categories: device classification in IoT, network traffic analysis and time-series data classification. 

\subsection{Device classification in IoT}
Yair Meidan et.al \cite{third} approach the IoT device classification from the network security perspective with the goal of checking whether a device is on a whitelist of devices that are approved to be connected to the network. They exclusively focused on features extracted from TCP sessions and use the Random Forest ML algorithm as their classifier. Arunan Sivanathan et.al \cite{second} develop an IoT device classification technique to identify an unique IoT device by examining its network traffic. They extracted 12 attributes related to time, packet length, protocols and pass them to a Random Forest classifier. While they achieve good classification accuracy, their method requires that the classifier must be trained using network traffic from each and every device under consideration. This is impractical given the large number of IoT devices on the market today. Yao et. al \cite{eighteen} proposed a graph-based object classification method, where a correlational graph of objects is established via random walk with restart based approach. A probabilistic feature-rich model is then proposed to categorize the heterogeneous objects under a multi-label classification scheme. Markus Miettinen et.al \cite{nineteen} also conduct device-type identification research for security enforcement in IoT. Their goal is to identify the device-types of new IoT devices when they are introduced into the network, so their method could only identify devices during the initial setup period. Moreover, the devices in one type are required to have the same firmware/hardware versions. This pre-condition is not practical in real-world.


\subsection{Network Traffic Analysis}
Our work also has some similarities with the broader topic of network traffic analysis which has been widely studied in recent years \cite{seven}. As defined in \cite{eight}, ``network traffic analysis studies inferential methods which take the network traces of a group of devices (from a few to thousands) as input, and give information about those devices, their users, their apps, or the traffic itself as output". Network traffic contains a lot of useful information about the type of devices, users and applications being used. Thus analysis of this traffic is useful for a number of applications such as network intrusion detection \cite{nine}, app identification \cite{ten}, in-app service usage classification \cite{fifth}\cite{eleven}, and user fingerprinting \cite{twelve}. 

Depending on the application scenario, different features are extracted from the raw network traffic. For example, traditional methods for in-app service usage classification analyze the TCP (or UDP) port numbers and IP addresses of IP packets to estimate the usage types. A more recent method uses features such as packet length statistics and time interval for every two consecutive packets \cite{fourth}. For IoT intrusion detection, researchers consider features like signal strength and packet types such as TCP SYN or TCP ACK instead \cite{thirteen}. These features and feature extraction methods are quite instructive for our task.

\subsection{Time-series Data Classification}
Our work also has a close relationship with time series classification because network traffic contains a sequence of packets with monotonically increasing time stamps and thus represents a time series. Time series data represents a collection of values obtained from sequential measurements over time \cite{fourteen}. In fact, many measurements in the real world are performed over time across a wide ranging scientific fields where classification is one of an important data mining tasks. Examples of such fields include Electroencephalography(EEG) signal in Brain-Computer Interface, motion sensor data in activity recognition, financial data in the stock market and computer log data. 

Many time series classification methods have been proposed over the past years. Geurts et.al \cite{fifteen} classify time series data based on a piece-wise representation and the resulting method is not robust to outliers. Xi et.al \cite{sixteen} use 1-NN classification algorithm with DTW as their classifier, but repeated DTW computations impair affect the computing speed significantly. Deep neural networks have been applied to time series classification as well. For example, Zachary et.al \cite{first} use LSTM recurrent neural networks to classify multivariate pediatric intensive care unit (PICU) time series to diagnose diseases. Xiang et.al \cite{seventeen} use a seven-layer neural network to recognize user intents to enable mind-controlled robots. Both studies outperform the state-of-the-art methods and demonstrate the feasibility of deep neural networks, especially recurrent neural network, in time series classification. These findings inspired us to explore the use of deep neural networks for the device classifcation problem at hand.


\section{The Proposed Approach}
\begin{figure}[!h]
\setlength{\abovecaptionskip}{0.cm}
\setlength{\belowcaptionskip}{-0.cm}
\centering
\includegraphics[width = 1.0\linewidth]{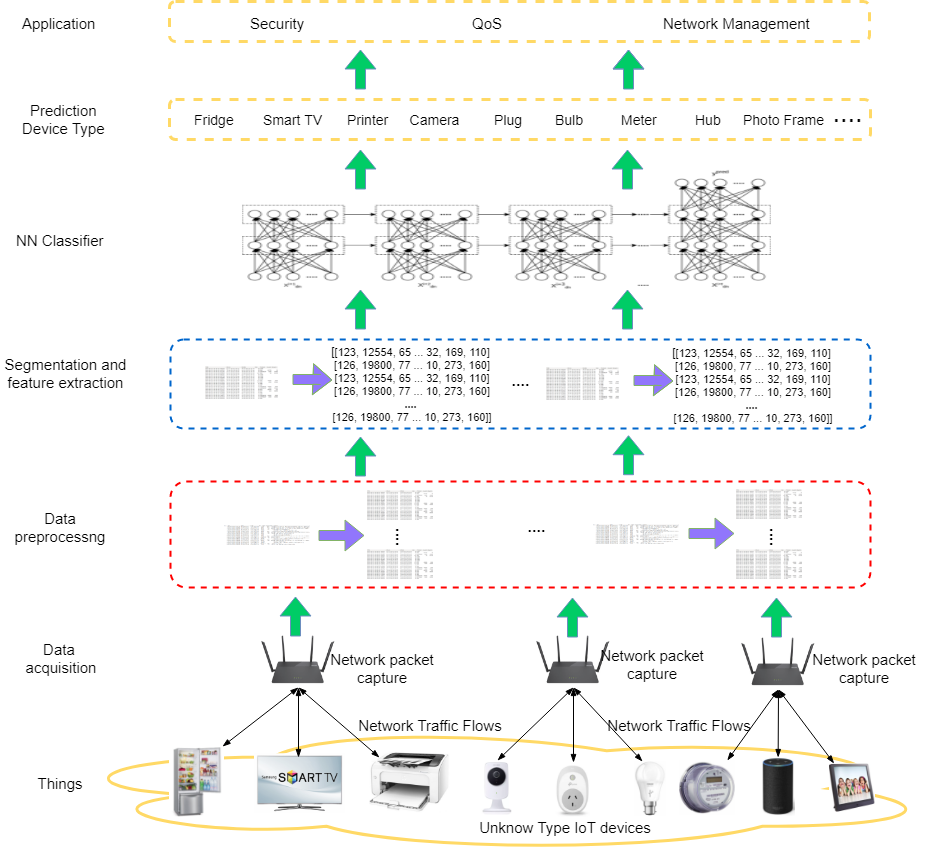}
\caption[]{Proposed Approach for Automatic Cross-Device identification}
\label{fig:workflow}
\end{figure}

In this section, we first present an overview of the proposed cross-device identification approach and then elaborate on the specific methodology. Fig.\ref{fig:workflow} depicts the approach which is comprised of three main components including: (i) Network Traffic Acquisition and Preprocessing (ii) Segmentation and Feature Extraction and (iii) Device Type Classification.

\subsection{Network Traffic Acquisition and Preprocessing}
Once connected to the network, IoT devices will generate traffic (incoming and outgoing) depending on certain configuration functions and application services. These packets include network configuration traffic (e.g., NTP, DNS), routine communication between the device and back-end server (e.g., keep alive messages) and traffic generated due to user interaction (e.g., the user initiating an Amazon Echo query). While different devices in the network could use different protocols and transmit data for different purposes, an overwhelming majority of this traffic uses TCP/IP protocols\cite{second}. 

\begin{figure*}
\centering
\includegraphics[width = 0.8\linewidth]{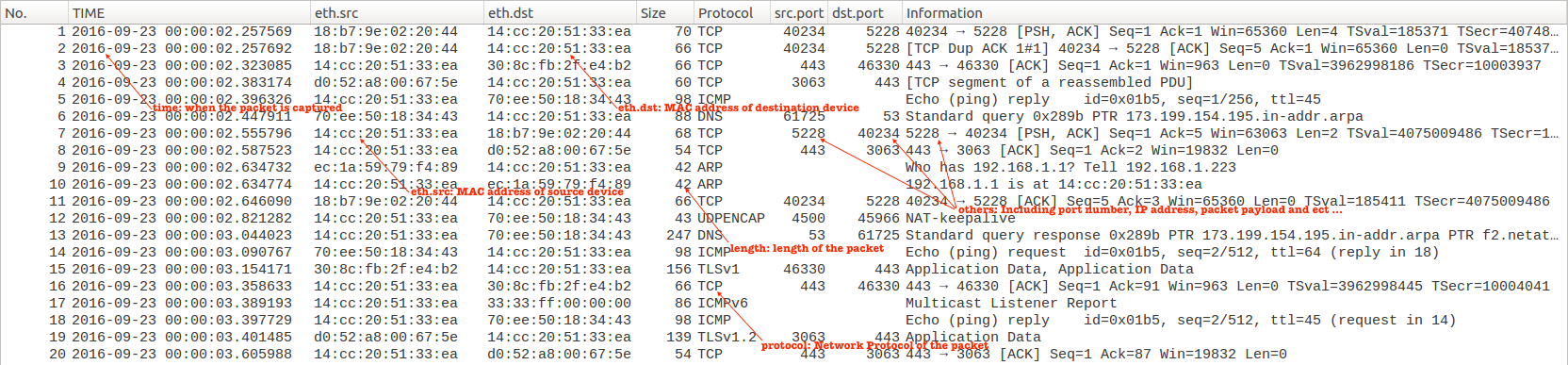}
\caption[]{An example of the captured traffic flow}
\label{packet_records}
\end{figure*}

As noted in Section 2.3, network traffic can be considered to be time series data and contains useful information about user habits, devices, and network status. We use network packet analyzers such as Wireshark\footnote{https://www.wireshark.org/} and tcpdump\footnote{https://www.tcpdump.org/} to capture network traffic packets. Packet analyzers running in the router or gateway can see all the device incoming and outgoing traffic and produce corresponding records. Each record contains all the information within that packet, from MAC layer to application layer. Due to the wide deployment of security protocols such as Secure Sockets Layer (SSL), Transport Layer Security (TLS) and the privacy protection policies of governments, only packet header could be utilized to make device classification. Fig.\ref{packet_records} gives an example of the captured traffic flow that contains a record of 20 packets and related explanations.

The traffic flow could be described as a sequence:
\begin{equation}
S = \{P^{1},P^{2},P^{3},P^{4},P^{5},\dots,P^{j},\dots \}
\end{equation}
where $P^{j}$ represents the information recorded for the $j-th$ packet. Each packet record $P^j$ contains all information in corresponding packet and is stored as: 
\begin{equation}
P^j = \{ t^j, length^j, protocol^j, eth.src^j, eth.dst^j, others^j \}
\end{equation}
where $t^j$ represents the approximate time when the packet is sent out or received, $length^j$ represents the packet length, $protocol^j$ represents the network protocol that this packet is using, $eth.src^j$ and $eth.dst^j$ represents the MAC address of source device and destination device respectively, $others^j$ represents all the other information captured but not used in our work. Here we note that these packets are recorded in time order $ t^1 < t^2 < \dots < t^j < \dots $.

Considering a network of N devices represented as $d_1, d_2, d_3, \dots, d_n, \dots (1\leq n \leq N)$, the traffic flow is described as:
\begin{equation}
S = \{P^{1}_{d_1},P^{1}_{d_2},P^{1}_{d_3},P^{2}_{d_1},P^{2}_{d_3},\dots,P^{i}_{d_n},\dots \}
\end{equation}

where $P^{i}_{d_n}$ means the $i_{th}$ packet of device $d_n$. In the data pre-processing part, we need to extract device specific sequences for each device and filter out information from $P^j$ that is not useful. Each device can be uniquely identified  based on the MAC address in P (eth.src or etht.dest depending on the direction of the traffic). Subsequently, device specific packet streams can be separated as follows:

\begin{equation}
S_{d_n} = \{P^{1}_{d_n},P^{2}_{d_n},P^{3}_{d_n},\dots \}
\end{equation}

\subsection{Segmentation and Feature Extraction}
As noted in Section 3.1, several pieces of information can be recorded from the network traffic including packet length, timestamp, protocol, etc. Many useful features could further be extracted from these information. 

\begin{figure}
\setlength{\abovecaptionskip}{0.cm}
\setlength{\belowcaptionskip}{-0.cm}
\vspace{0.1cm}
\centering
\includegraphics[width = \linewidth]{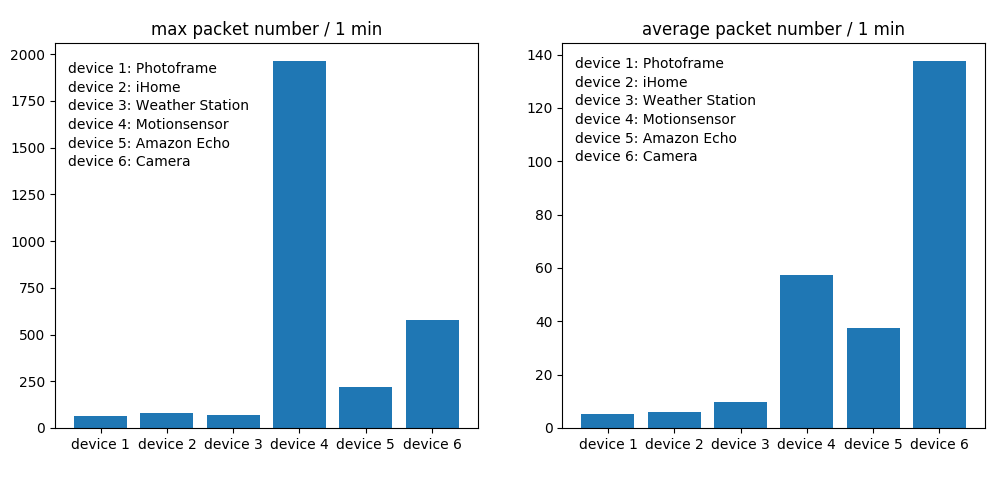}
\caption[]{from left to right: (a)maximum packets number produced by six common IoT devices in one minute; (b)average packets number produced by six common IoT devices in one minute;}
\label{packet number}
\end{figure}

Since each device generates a large amount of traffic, segmentation of the traffic is necessary before proceeding with feature extraction. Fig.\ref{packet number} illustrates the maximum and average traffic load attributed to different IoT devices. As can be observed the traffic intensity varies from device to device and also for different time scales. For example, a motion sensor generates near 1900 packets per minute at most and a camera generates up to 140 packets per minute in average. While these packets contain useful information, they are liable to be redundant and superfluous. Though deep neural networks are good at finding patterns hidden in raw data, processing each packet individually would be both time-consuming and computationally exhausting. Thus we segment raw traffic flows first before extracting features.

\subsubsection{Segmentation:}
In our approach, we segment the traffic flow into sub-traffic flows of a fixed time interval $T$ with the help of timestamps in records. For example, the traffic flow from device $d_1$ can be expressed as multiple segments of duration $T$ as follows:

\begin{equation}
S_{d_1} = \{sub^{0-T}_{d_1}, sub^{T-2T}_{d_1}, sub^{2T-3T}_{d_1}, \dots, sub^{iT-(i+1)T}_{d_1}, \dots \}
\end{equation}
where $sub^{iT-(i+1)T}_{d_1}$ represents all the records of device $d_1$ in the time period of $iT$ to $(i+1)T$ and could be written as:
\begin{equation}
\setlength{\abovedisplayskip}{3pt}
\setlength{\belowdisplayskip}{3pt}
sub^{iT-(i+1)T}_{d_1} = \{P^{l}_{d_n},P^{l+1}_{d_n},P^{l+2}_{d_n},\dots, P^{l+m}_{d_n}, \dots \}
\end{equation}
Segmentation condition is $iT \leq t^{l+m}_{d_n} \leq (i+1)T$.

\subsubsection{Feature Extraction:}
A large number of features could be extracted from different perspectives from the raw data of a segmented sub-traffic flow $sub^{iT-(i+1)T}_{d_1}$. Our approach extracts features mainly from the perspective of traffic volume, packet length, network protocols and direction of traffic (i.e. incoming vs outgoing). 
According to the packet protocol, we divide $sub^{iT-(i+1)T}_{d_1}$ into two categories: user packets and control packets. User packets include user data and device-server communication packets. A packet $P^j$ is classified as a user packet if $protocol^j$ is either TCP, UDP, HTTP or other high layer protocols. Control packets mainly are supporting functional protocol packets such as ICMP packets, ARP packets, DNS packet and NTP packets. According to packet directions, $sub^{iT-(i+1)T}_{d_1}$ could also be divided into two categories: received packets and transmitted packets.
The extracted features could be divided into following types.
\begin{itemize}
    
    \item [\textbullet]features related to the number of packets: This category includes the number of total packets, user packets, control packets, received packets, transmitted packets and packet counts for different protocols such as DNS, ARP, NTP. As noted in Section 3.1, different types of devices generate varying traffic. Thus, some devices result in more informative features than others. 
    
    \item [\textbullet]packet length statistics: As a segmented sub-traffic flow contains many packets of different lengths, it is vital to explore the statistics of these packet lengths. Therefore, we extract the first order and the second order statistics features including maximum, minimum, mean, sum, standard deviation, variance, skewness, and kurtosis. We also extract these statistics features from the aforementioned four categories of  $sub^{iT-(i+1)T}_{d_1}$.
    
    \item [\textbullet] protocol related features: Features of this type include the count of different types of protocol packets contained within a segment $sub^{iT-(i+1)T}_{d_1}$.
    
\end{itemize}


As a result of feature extraction, $sub^{iT-(i+1)T}_{d_1}$ could be represented as a features vector $x^i_{d_1}$, and the traffic flow of device $d_1$ is represented by $S_{d_1} = \{ x^0_{d_1}, x^1_{d_1}, x^2_{d_1}, \dots, x^i_{d_1}, \dots \}$.

\subsection{Device Classification}

\begin{figure*}
\centering
\includegraphics[width = \linewidth]{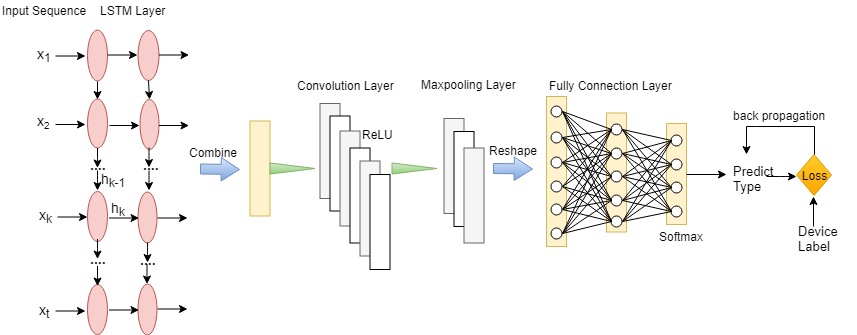}
\caption[]{Proposed LSTM-CNN cascade device classification model}
\label{model}
\end{figure*}

In order to classify different IoT devices, we propose an end-to-end classification method based on deep learning algorithms. The model is shown in Figure \ref{model}. Suppose $d_n$ belongs to type $y$. To classify devices with network traffic flows, the model needs to be trained with a traffic flow $S_{d_n}$ to generate a device type prediction ${y_{pred}}$. Inputs of the model $\{ x^{i+1}_{d_n}, x^{i+2}_{d_n}, x^{i+3}_{d_n}, \dots, x^{i+t}_{d_n} \}$ are a part of the extracted feature sequence $\{ x^0_{d_n}, x^1_{d_n}, x^2_{d_n}, \dots, x^i_{d_n}, \dots \}$, where ``t" is time window size and could be defined according to data. Output is the predicted device type ${y_{pred}}$.

$\bm{LSTM Layer.}$ The inputs are fed into two LSTM layers at first to capture the temporal relationship of network traffic. LSTM is a prominent variation of Recurrent Neural Networks (RNN) specially designed for processing sequential data. It has been shown to be very effective for many applications ranging from speech recognition, handwriting recognition, and anomaly detection. For a basic LSTM cell, the inputs include two parts: $\bm{x_k}$ and $\bm{h_{k-1}}$, where $\bm{h_{k-1}}$ is the output of last cell in the same layer. The computation of a LSTM cell is defined by the following equations \cite{first}:

$$ \bm{g}^{(k)} = tanh(W^{\textup{gx}} \bm{x_k} + W^{\textup{gh}} \bm{h_(k-1)} + \bm{b}^{\textup{g}})$$
$$ \bm{i}^{(k)} = \sigma(W^{\textup{ix}} \bm{x_k} + W^{\textup{ih}} \bm{h_(k-1)} + \bm{b}^{\textup{i}})$$
$$ \bm{f}^{(k)} = \sigma(W^{\textup{fx}} \bm{x_k} + W^{\textup{fh}} \bm{h_(k-1)} + \bm{b}^{\textup{f}})$$
$$ \bm{o}^{(k)} = \sigma(W^{\textup{ox}} \bm{x_k} + W^{\textup{oh}} \bm{h_(k-1)} + \bm{b}^{\textup{o}})$$
$$ \bm{s}^{(k)} = \bm{g}^{(k)} \odot \bm{i}^{(k)} + \bm{s}^{(k-1)} \odot \bm{f}^{(k)}$$
$$ \bm{h}^{(k)} = tanh(\bm{s}^{(k)}) \odot \bm{o}^{(k)}$$
where $W^{\textup{gx, gh, ix, ih, fx, fh, ox, oh}}$ and $b^{\textup{g, i, f, o}}$ are parameters to be learned, $\bm{h}^{(t)}$ is the output of LSTM cell. For multi-layer LSTM, $\bm{h}^{(t)}$ is also the input of next layer.

$\bm{Convolution Layer.}$ Outputs of LSTM layer are t vectors. We concatenate them as columns to form a 2-D vector and feed the 2-D vector into the convolution layer, which is a special type of constrained feed-forward neural networks. The convolution layer use multiple filters go through the 2-D vector. Next, the outputs of these filters are passed to linear or non-linear activation functions such as Rectified Linear Unit (ReLU) to form the output feature maps. 

$\bm{Maxpooling Layer.}$ The output of convolution layer is then fed into maxpooling layer directly. The maxpooling layer will reduce the dimension of inputs by only selecting the maximum value from n*n features, where n*n is the maxpooling filter size.

$\bm{Fully Connection Layer.}$ After the maxpooling layer, data is reshaped to a vector again and passed into a fully connection layer with dropout operation before feeding into the output layer. The dropout operation helps prevent over-fitting and achieve better generalization. In the output layer, softmax function is chosen as the active function to calculate the probabilities of different classes. The class with the highest probability will be final prediction for inputs.

\section{Experimental Evaluation}
We use real-world data collected from IoT devices to evaluate the feasibility of our approach. We first provide an overview of the dataset. Next we present the performance results and also compare our method to several strong baselines. We also present a sensitivity analysis to evaluate the impact of different parameters on the results. This section concludes with a discussion.  

\subsection{Dataset}
\begin{figure}
\setlength{\abovecaptionskip}{0.cm}
\setlength{\belowcaptionskip}{-0.cm}
\centering
\includegraphics[width = \linewidth]{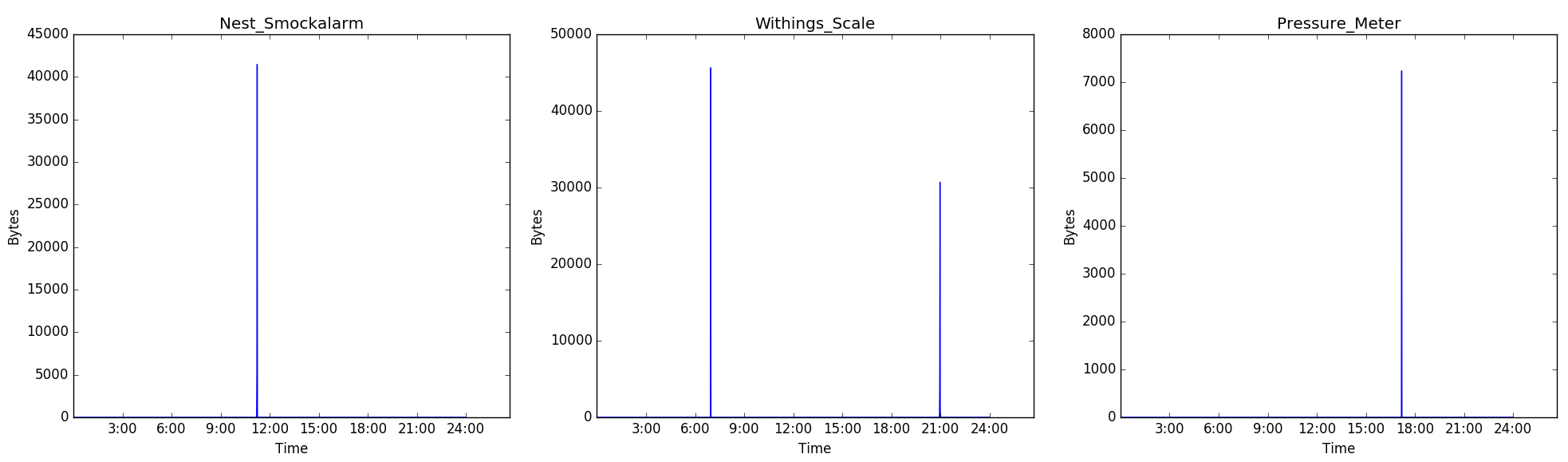}
\caption[]{from left to right: daily activity of (a)NEST protect smoke alarm; (b)Withings smart scale; (c)Blipcare blood pressure meter}
\label{sensor data}
\end{figure}

The original data is collected by Arunan et.al.\cite{second} in a IoT campus environment over 3 weeks. It contains traffic from 21 devices which were classified into 7 categories. However after examining the data closely, we found some of the devices produce network traffic only in limited time which cannot truly be qualified as time series data. Fig.\ref{sensor data} shows daily network activities of 3 such devices which were thus excluded. Subsequently, we found that certain categories only contain a single device. Since our objective is to test cross-device classification, we also excluded these devices from our analysis. In the end, we are left with 15 devices belonging to 4 categories. These devices are summarized in Table.\ref{table:device}. 

\begin{table}[!h]
\setlength{\abovecaptionskip}{0.cm}
\setlength{\belowcaptionskip}{-0.cm}
\caption{Device List in each category}
\begin{center}
\begin{tabular}{| >{\centering\arraybackslash}m{0.9in} |>{\centering\arraybackslash}m{1.3in} |>{\centering\arraybackslash}m{0.4in}  | >{\centering\arraybackslash}m{0.2in} | }
\hline
Category             & Device  &Raw Item  & Label \\
\hline
\multirow{2}*{Hubs}  & Amazon Echo  & 1310K                   & \multirow{2}*{1}     \\
                     & Smart Things  & 553K                 &       \\
\hline
\multirow{3}*{Electronics}          & Triby Speaker   &194K                & \multirow{3}*{2}     \\
                     & PIX-STAR Photo-frame  &67K           &       \\
                     & HP-Printer      &241K                &       \\
\hline
\multirow{6}*{Cameras}              & Netatmo Welcome    &659K             & \multirow{6}*{3}     \\
                     & Withing Smart Monitor  &655K    &       \\
                     & Samsung SmartCam    &1261K            &       \\
                     & TPLink Day/Night Cloud Cam  &310K   &       \\
                     & Dropcam       &4122K                  &       \\
                     & Insteon Camera    &608K              &       \\
\hline
\multirow{4}*{Switches\&Triggers} & Belkin Wemo Switch   & 1092K           & \multirow{4}*{4}     \\
                     & TP-Link Smart Plug   & 39K           &       \\
                     & iHome   & 50K                        &       \\
                     & Belkin Wemo motion sensor   & 1434K    &       \\
\hline
\end{tabular}
\end{center}
\label{table:device}
\end{table}

\subsection{Overall Results and Comparison}
In this section, we present the overall classification results of our method and compare our approach with several baselines. To test the performance of classifying an unknown device type, the traffic stream of this device should be excluded from the training data. We chose the devices in each category for training and used the remaining devices from that category for testing. More specifically, we pick Amazon Echo from Hubs, Belkin Wemo Switch, TP-Link Smart Plug from Switches\&Triggers, Pix photo frame from Electronics and Withing Smart Baby Monitor, Netatmo Welcome, and Samsung Smart Camera from Cameras, and use their data as the training data. Data from the remaining devices are used as testing data. 

We set the segmentation time interval $T$ to 5 minutes, as well as normalize and shuffle the extracted features before feeding them into our classification model. After some preliminary tests of different feature combinations, we choose 6 most discriminating features: user packet number, user packet length average, user packet length peak, control packet number, control packet average, control packet peak. The time window size (described in sec 3.3) of our model is set to 6 and there is a 50\% overlap (i.e. overlap is 3 when time window size is 6) between successive windows. In convolution layer, we use 32 2*2 filters with 1*1 stride. In maxplling layer, both filter size and stride are 2*2. Besides, our classification model has other three hyper-parameters:dropout probability, learning rate, and coefficient $\lambda$ for L2 normalization. After hyper-parameter tuning, we set their value as 0.8, 0.05 and 0.01, respectively. 

\begin{figure}
\setlength{\abovecaptionskip}{0.cm}
\setlength{\belowcaptionskip}{-0.cm}
\centering
\includegraphics[width = 0.6\linewidth]{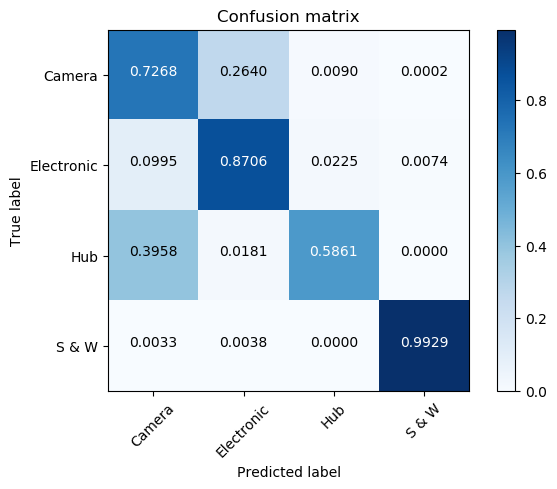}
\caption[]{The confusion matrix of device classification}
\label{confusion matrix}
\end{figure}

Under these setting, we repeat the experiment for five times and get the average accuracy of 74.8\% and best accuracy of 80.1\%. Figure \ref{confusion matrix} shows the confusion matrix of the best classification results. Considering that we only use half of the devices within each category for training the model, this result is fairly promising. 

\subsubsection{Baseline comparison}
Since the authors in  \cite{second} and \cite{third} use data from the same devices as the training data and testing data, comparing our method with theirs is unfair. Therefore, in this part, we compare our method with several widely used classification techniques. 
\begin{itemize}
    \item[\textbullet] Support Vector Machine(SVM): SVM algorithm maps original data to high-dimensional space and constructs a hyperplane that maximizes the separation between two classes.
    \item[\textbullet] k-Nearest Neighbor(kNN): kNN is one of the simplest classification algorithms but works well in practice. It predicts by searching the training set for the k most similar instances. 
    \item[\textbullet] Decision Tree: Decision tree uses a flowchart-like structure and makes a decision in each internal node by attributes. Leaf nodes represent the classification label.  
    \item[\textbullet] Random Forest(RF): A random forest is a collection of decision trees.
    \item[\textbullet] AdaBoost: AdaBoost, or adaptive boosting, is a successful boosting algorithm.
    \item[\textbullet] Linear Discriminant Analysis(LDA): LDA aims at finding a linear representation of features that could separate different objects.  
    \item[\textbullet] Quadratic Discriminant Analysis(QDA): QDA is more general than LDA; it assumes that the data for each class are normally distributed. 
    \item[\textbullet] Multilayer perceptron(MLP): MLP is a typical class of feedforward neural network. It has multiple layers and non-linear activation function. Through elaborate design, MLP could perform well in many tasks.
    \item[\textbullet] Convolutional Neural Network(CNN): CNN is another important category of neural networks. It's hidden layers typically consist of convolutional layers and pooling layers. CNN only needs little pre-processing or even no pre-processing of the raw data.  
\end{itemize}

Our implementation uses SVM adopting the One-VS-Rest strategy as the classifier and LinearSVC as the estimator. For the parameters, we set the number of neighbors to 10 in kNN, the number of estimators in RF to 300, the max depth of decision tree to 12, the number of estimators of AdaBoost to 50, and the learning rate to 0.3. All other parameters are set to default values. Table \ref{table:baseline} shows the final classification results comparison. Our approach significantly outperforms all of the other approaches, demonstrating the effectiveness of LSTM-CNN cascade model for automatic device classification. The results also suggest the feasibility of automatic device classification with network traffic flows.

\begin{table}[!]
\caption{Baseline comparison of our approach with other methods }
\centering
\begin{tabular}{c c c}
\hline
Index & Methods       & Accuracy(\%) \\
\hline
1     & SVM           & 58.5     \\
2     & RF            & 30.1     \\
3     & KNN(k=10)           & 27.6    \\
4     & Decision Tree & 46.4     \\
5     & AdaBoost           & 48.5    \\
6     & LDA           & 49.4    \\
7     & QDA           & 52.4    \\
8     & MLP           & 52.1      \\
9     & CNN           & 56.3     \\
10    & LSTM          & 65.4     \\
11    & Ours          & \textbf{74.8}    \\
\hline
\end{tabular}
\label{table:baseline}
\end{table}

\subsection{Impact of Different Parameter Settings}
We have evaluated the impact of the segmentation time interval and time window size on the performance of our proposed approach. Each test is conducted five times and the average result is calculated to ensure the reliability of our data. 
\subsubsection{Segmentation time interval}
As described in section 3.2, our features are extracted from segmented traffic flows by subdiving the traffic stream into distinct windows of duration $T$. On the one hand, an excessively large segmentation time interval may result in very similar traffic characteristics over each segment which would result in very little variations in the feature vector of a device. It could also lead to fewer data samples for training and testing, and consequently, has a negative impact on neural networks. On the other hand, when the segmentation time interval is too small, the features in a time period would remain steady and become redundant in reflecting the patterns.

\begin{figure*}[htbp]
\centering
\begin{minipage}[t]{0.3\textwidth}
\centering
\includegraphics[width=5.5cm]{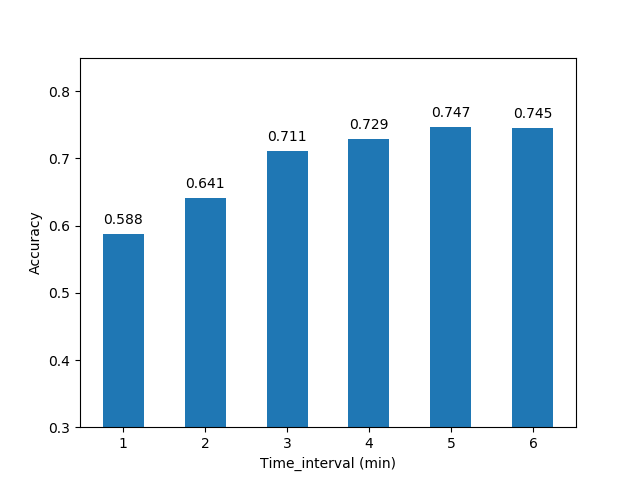}
\caption{Classification results under different segmentation time interval}
\label{segementation interval}
\end{minipage}
\begin{minipage}[t]{0.3\textwidth}
\centering
\includegraphics[width=5.5cm]{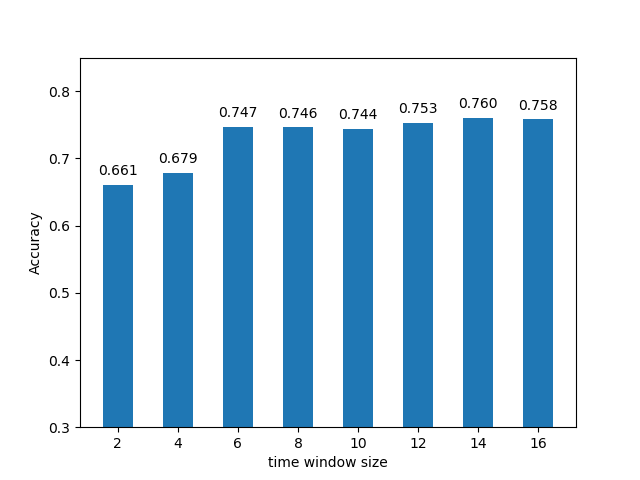}
\caption{Classification results under different time window size}
\label{time window}
\end{minipage}
\begin{minipage}[t]{0.3\textwidth}
\centering
\includegraphics[width=5.5cm]{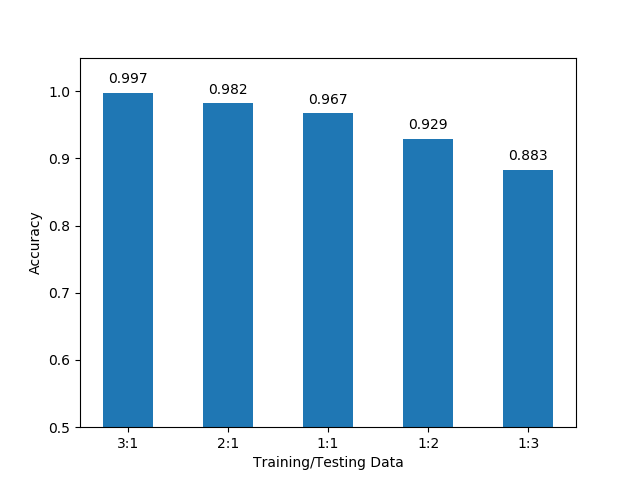}
\caption{The overall accuracy of device classification over different ratio of training data and testing data}
\label{result:ratio}
\end{minipage}
\end{figure*}

Figure \ref{segementation interval} shows the influence of segmentation time interval on classification accuracy. In these tests, all time-window size is set to \textbf{6} while learning rate and $\lambda$ are tuned for each time interval. Figure \ref{segementation interval}, we could see that when the segmentation time interval is too small, such as 1 minute, the classification results are significantly lower than larger segmentation time intervals.

\subsubsection{LSTM time window size}
Time window size is another important parameter for investigation. Figure \ref{time window} presents the classification accuracy under different time window size when the segmentation time interval is \textbf{5} minutes. The learning rate and $\lambda$ also are tuned for each time window size. For time series data such as network traffic flows, a larger time windows size always means a longer period of network traffics as input. While a larger time window could learn patterns from traffic flows more accurately, it needs more inputs and, in turn, needs data collection over an extended period of time. The time window size would not influence the accuracy significantly when it is larger than 8. 

\subsection{Binary Classification Results}
In the previous sections, we present the classification results for a fixed ratio of training data. In this section, we evaluate the performance of our method under different ratios of training and testing data. Since there is limited data available for the devices that belong to Hubs and Electronics categories, we only consider devices from the remaining two categories, i.e. Switches\&Triggers and Cameras, which means this experiment is a two-type
classification.

Figure \ref{result:ratio} shows the overall accuracy of device classification over different ratios of training data and testing data for our proposed method. The results show our method can achieve a very high accuracy (99.7\%) when 75\% of the data is used for training. Even when we only use 25 percent as the training data, our method can still achieve an overall accuracy of 88.3. 

\subsection{Discussion}
In this section, we present our insights on model analysis and IoT data acquisition in practical deployment.

Our proposed approach is a generic method for IoT device classification. Our experimental results in Section 4 demonstrate that our approach is feasible for automatic classification of new IoT devices by only analysing their network traffic streams, which are generally easily obtainable. Before applying this model to various IoT application scenarios, we would like to illustrate some key findings which have significant implications on evaluating our approach and potentially other device classification approaches using real-world IoT data.  
During our empirical evaluations, we observed that the performance degrades with the increase in the number of device types. When using 50 percent data as training data, the classification accuracy for two-class classification was 96.7\%. But when expanding to the four-class classification scenario, the accuracy dropped to 74.8\%. The confusion matrix of a four device classification test (shown in Table \ref{confusion matrix}) shows that our approach fails to distinguish Hubs accurately. The main reason can be attributed to the rather limited information available in the small scale dataset that we have used. More specifically, there are only two devices in Hubs category, which means we could only use one device's data as the training data. However, a single device may not necessarily provide sufficient information that is representative of the entire category. Even though the dataset we are using contains millions of raw packets from 15 IoT devices in 19 days, which could arguably be considered a large dataset from the perspective of traditional network traffic analysis, it is still relatively small for our purpose.

Based on the above analysis, we believe the richness of dataset is more crucial than the scale of the dataset for IoT device identification, especially for the multi-class scenario. To increase the diversity and richness of IoT dataset for better practical use, more devices in each category, along with a broader range of devices should be considered during the stage of data acquisition.  



\section{Conclusion}
In this paper, we propose to use network traffic flows to classify new and unseen IoT devices automatically. After preprocessing, segmentation, and feature extraction, we use a LSTM-CNN cascade model to conduct cross-device classification. We evaluate our approach by classifying 15 IoT devices into four types with real-world collected network traffic data and achieve an accuracy of 74.8\%. This result outperforms a series of widely used classification algorithms. Besides, we also examine the performance of our method under different ratios of the training data and get an accuracy of 99.7\% when using 75 percent data as training data. Although there is still room for improvement, our work successfully demonstrates the possibility to automatically classify IoT devices based on their network traffic flows. While traffic data can be readily available within an organization, our method can be deployed easily with low cost to enable a more intelligent IoT network. Our future focus would be on building a testbed that contains more IoT devices and user interactions and further improving the classification accuracy. Based on the enhanced testbed, we will also evaluate the efficiency of our approach under different network scale and traffic volume settings to test the applicability in high speed IoT scenarios.

\end{document}